# Towards Expert-Based Speed–Precision Control in Early Simulator Training for Novice Surgeons

**Birgitta Dresp-Langley**

CNRS UMR 7357 ICube Lab, Strasbourg University, Strasbourg 67000, France; birgitta.dresp@icube.unistra.fr; Tel.: +33-388-119-117

**Abstract:** Simulator training for image-guided surgical interventions would benefit from intelligent systems that detect the evolution of task performance, and take control of individual speed–precision strategies by providing effective automatic performance feedback. At the earliest training stages, novices frequently focus on getting faster at the task. This may, as shown here, compromise the evolution of their precision scores, sometimes irreparably, if it is not controlled for as early as possible. Artificial intelligence could help make sure that a trainee reaches her/his optimal individual speed–accuracy trade-off by monitoring individual performance criteria, detecting critical trends at any given moment in time, and alerting the trainee as early as necessary when to slow down and focus on precision, or when to focus on getting faster. It is suggested that, for effective benchmarking, individual training statistics of novices are compared with the statistics of an expert surgeon. The speed–accuracy functions of novices trained in a large number of experimental sessions reveal differences in individual speed–precision strategies, and clarify why such strategies should be automatically detected and controlled for before further training on specific surgical task models, or clinical models, may be envisaged. How expert benchmark statistics may be exploited for automatic performance control is explained.

**Keywords:** surgical simulator training; individual performance trend; speed–accuracy function; automatic detection; performance feedback

## 1. Introduction

Technological development and pressure towards a reduction in time available for learning has radically changed the traditional apprenticeship model of surgical training, where simulation now offers the opportunity for repeated practice in safe and controlled environments. The complexity and reliability of commercially available simulators varies considerably, and selecting an appropriate simulator for surgical skill training is in itself a challenge. Simulators for specific surgical skills are generally tested for the highest validity level [1], that of predictive validity, ensuring that assessments of performance in the specific simulator task are likely to predict future performance of the trainee in an equivalent task in the clinical context (animal, patient). Only a certain percentage of surgical simulators provide some kind of performance feedback to the trainee. The feedback systems as such are generally not validated. In other words, whether the feedback given during training is actually truly useful to a novice is not known. Ideally, within a surgical curriculum, trainees should have dedicated time for simulation-based training with appropriate performance monitoring through effective feedback systems, as the main advantage of computer simulators for surgical training is the opportunity they afford for independent learning. Yet, if the simulator does not provide relevant and truly useful instructional feedback to the user, then instructors need to be present to supervise and tutor the trainee. The way feedback is delivered to the operator, and the amount of feedback he/she has to process and integrate, represents an important challenge in the development of automatic systems for simulator training assistance [2,3]. If relevant information is not delivered effectively, either by drowning essential feedback parameters in a large amount of unnecessary ones, or by not providing truly useful feedback that will help the trainee gain proficiency at the task at hand, then operators will not be able to understand the message given by the system, and the latter will have failed its purpose altogether. Guidelines for user interface design and feedback procedures in simulator training contexts do not yet exist, but they can and should be worked out and tested [2,3].

Relevant performance metrics [4–10] are essential to the development of surgical simulator systems for optimal independent training. The presentation of such metrics to the user, in a way that boosts independent learning by producing a measurable skill improvement, is the most important aspect of an effective training system [11]. Metric-based simulation ensures that training sessions are more than just simulated clinical procedures and gets rid of subjectivity in evaluating skill evolution; there is no ambiguity about the progress of training. Benchmarking individual levels of proficiency on the performance levels of experts in a validated, metric-based simulation system has well-established intrinsic face validity [11], and appears a better approach than benchmarking on performance concepts based on expert consensus, for example. Building expert performance in terms of benchmark metrics into simulator training programs provides a sound basis for automatic skill assessment. Benchmarking ensures that the "pass" level is defined by realistic criteria, set directly by the proficiency levels of individuals who are highly experienced at performing the clinical procedures that simulator training is aimed at preparing novices for [11–15].

Whether artificial intelligence (AI) can help improve surgical simulator training is still by and large an open question. AI provides well-suited concepts for knowledge implementation, automatic feedback procedures, and the exploitation of prior (learned) benchmark knowledge; building such procedures into simulator training could have clear benefits, especially at early stages of training. Early-stage "dry-lab" training is offered to large numbers of novices, often on experimentally developed simulators, and supervision of the training programs by one or two experts may not be the best way of ensuring optimal training. Automatic control procedures [13] using metric-based benchmark criteria and statistically driven performance comparisons, with trial-by-trial feedback at any given moment in time, may prove the better alternative. The goal of early simulator training is to help the largest possible number of registered individuals reach optimal performance as swiftly as possible [15–23], and therefore requires systems of skill monitoring aimed at tutoring each and every single individual rather than merely assessing end-of-session performance status, or differences between users after training.

This concept paper here, in the light of the analyses that will follow, suggests an early simulator training model for automatic skill evolution on the basis of individual speed–precision data. These are exploited in comparison with benchmark statistics from an expert surgeon in the context of an experimental simulator environment. The approach is based on a simple and universal psychophysical human performance model [24–31]. It allows individual strategies during motor learning to be told apart on the basis of individual speed–accuracy trade-off functions. Automatic performance control and feedback may be implemented at any step of the training procedure in the simplest possible way, by using criteria relative to the mean and standard deviation of individual performance. The goal here is not to validate a set of algorithms, or a specific AI procedure. We first need to clarify which kind of background knowledge, performance criteria, and benchmarking decisions would need to be implemented into an AI system. Choosing a code that will perform the suggested procedure is a side-issue, knowing that many different simulator tasks exist [13,14,16,22], and that the procedure suggested here could be implemented into any simulator system that provides objective metrics relative to task execution time and precision. Any such specific simulator may then be validated in light of an expert's benchmark performance statistics, as suggested below.

The experimental simulator on which the data here were generated brings to the fore meaningful differences between an expert surgeon and any novice with equivalent number of simulator training trials in the same camera view conditions, as will be made clear in the light of the analyses and discussion. It will be shown that individual speed–precision trade-off functions of complete novices, trained for a large number of simulator sessions, reveal different speed–precision strategies. The reasons why such strategies need to be detected, and if necessary, controlled for and modified as early as possible in simulator training, are then discussed. In the final step of the analysis, a principle of automatic expert benchmarking is brought forward to conceptualize key properties of an automatic procedure that (1) knows the expert surgeon's statistics (in-built benchmarks) relative to task execution time and task precision, (2) detects, and if necessary (3) controls for individual speed–precision strategies by comparing a trainee's performance statistics to the expert's benchmarks and, finally, (3) is able to provide appropriate user feedback when necessary. It is argued that such a system will enable any surgical trainee, without the intervention of a tutor and at the earliest stages of training, to attain the highest level of task precision he/she is capable of on the simulator. This should, ultimately, result in optimally trained "strategy-aware" individuals a surgical expert committee can select from to pick the best for further tuition. (Figure 5).

## 2. Materials and Methods

Data relative to the evolution of individual performance measures, relative to task speed and precision, were automatically monitored and recorded, using a specifically designed experimental simulator platform for image-based analysis of performance data relative to the time and precision of hand-tool movements in a five-step computer controlled pick-and-place task. The technical aspects of this platform, which was used in several experimental studies published elsewhere, are described in detail, with images and illustrations, in previous work [4–8].

*2.1. Research Ethics and Participants*

All experiments were conducted in conformity with the Helsinki Declaration relative to scientific experiments on human individuals, with the full approval of the ethics board of the corresponding author's host institution (CNRS) relative to non-invasive data collection from human individuals. All participants were volunteers and provided written informed consent. Their identity is not revealed. Data shown here are training sessions of fourteen novices with no experience in image-guided or other surgical procedures (absolute beginners). The data relating to the expert performance measures, shown for comparison, were recorded from single training sessions of a highly skilled expert endoscopic surgeon with more than 30 years of experience in image-guided surgery, but no training at all in the specific pick-and-place simulator task here.

*2.2. Camera Views*

2D and 3D camera views, shown on a 2D screen or in stereoscopic viewing using a head-mounted virtual reality device (OCULUS DK2, manufactured and distributed by Palmer Luckey Oculus VR, Menlo Park, California, USA), were generated through one or two 120° fisheye lens camera(s), fully adjustable in 360°. Snapshot views of critical simulator system parts are shown in Figure 1. The video input received from the camera(s) was processed by a DELL Precision T5810 model computer, equipped with an Intel Xeon CPU E5-1620 with 16 Gigabytes memory (RAM) capacity at 16 bits and an NVidia GForce GTX980 graphics card. The computer was connected to a high resolution color monitor (EIZO LCD "Color Edge CG275 W"), which communicates with the Color Navigator 5.4.5 interface for Windows. The color/grey levels of objects visualized on the screen could be matched to LAB or RGB color space, and the color coordinates for RGB triples could be retrieved from a look-up table at any moment in time. Task sessions and data generation were controlled by a program written in Python 2.7 for Windows, using the Open CV computer vision software library.

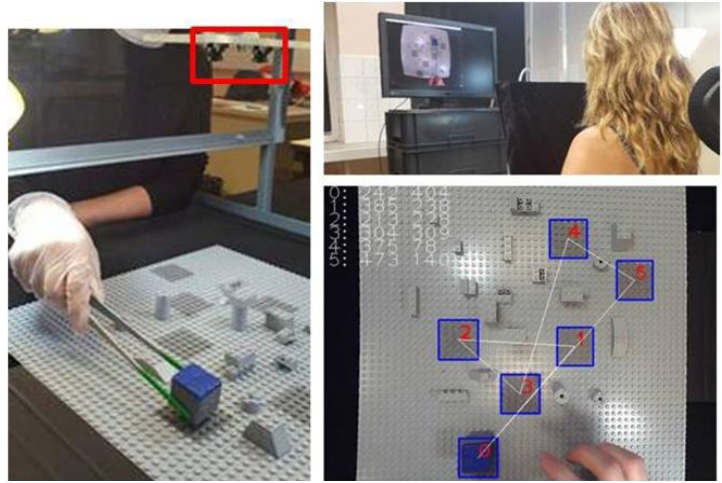

**Figure 1.** Snapshot views of critical parts of the experimental simulator. One or two cameras (highlighted by the red square in the image on the left) produced the 2D/3D screen views of the action field on the screen ahead of the trainee (**top image on the right**). Trainees practiced in sessions, with varying 2D/3D image views and monitor positions, to learn a five step pick and place task that involved placing a small blue foam cube as precisely as possible, using a forceps-like tool, on the center of five target areas in a specific order (**bottom image on the left**). Precision data in terms of "off-target" scores (in pixels) were computed using system coordinates relative to the target centers (in pixels) and surface-position coordinates relative to the blue foam cube (in pixels), tracked by the system (**bottom image on the right**).

*2.3. Five Step Pick-and-Place Task*

The task action field consisted of a classic square shaped (45 cm × 45 cm) light grey LEGO board available worldwide in the toy sections of large department stores. Six square (4.5 cm × 4.5 cm) target areas were painted on the board at various locations in a medium grey tint (acrylic). In between these target areas, small LEGO pieces of varying shapes and heights were placed to add a certain level of complexity to both the visual configuration and the task, and to reduce the likelihood of getting performance ceiling effects. In a five step pick-and-place task, a small (3 cm × 3 cm × 3 cm) cube made of very light plastic foam, but resistant to deformation in all directions, had to be placed on the target areas in a specific order (Figure 1). Trainees were generally right-handed, as those for whom data are shown here. They were instructed to position the foam cube, with their dominant hand and using a forceps-like tool, as precisely and swiftly as possible on the center of each target, in the right order. Data from fully completed trial sets only were recorded. A fully complete trial set consisted of a set of pick-and-place operations from target to target, in the right order and without dropping the object accidentally. Ten fully completed trial sequences were recorded in each training session. Different camera view conditions were carefully counterbalanced between individual sessions and between trainees, to avoid order effects during learning.

*2.4. Generation of Individual Training Data for Time and Precision*

For each single trial, the simulator system generated training data in terms of individual performance measures for task execution time (in milliseconds) and task precision (in pixels). For measures relative to time, the system counted and recorded the CPU time, from the moment the blue cube object was picked up by the participant to the time it was put on a given target. At the end of a given five step training trial, the cumulated value of these times was computed and recorded. The rate for image–time data collection was between 25–30 Hz, with an error margin of less than 40 milliseconds for any of the time estimates. For measures relative to precision, the system counted and recorded the number of blue object pixels at positions "off" the 3 cm × 3 cm central area of each of the five 4.5 cm × 4.5 cm target areas whenever the object was placed on a given target. At the end of a given five step training trial, the cumulated value of these numbers was computed and recorded. The standard errors of these positional estimates, determined in a calibration procedure, were below 10 pixels. Time and precision data were written to an excel file by the computer program, with labeled data columns for the different conditions, and stored in a directory for subsequent analyses.

**3. Results**

In our previous studies, simulator training data relative to time (in seconds) and precision (in pixels) of image-guided pick-and-place task performance were recorded from different study populations, including absolute beginners, novice surgeons without specific experience in image-guided simulator training, and expert surgeons with variable experience in image-guided surgery [2,3,5]. These previous studies were aimed at investigating the effects of different camera views, monitor positions and levels of expertise on simulator skill statistics. The analyses that will be shown here are motivated by different objectives, as clarified in the introduction above.

*3.1. Expert Benchmark Statistics*

In a first analysis, a comparison of individual statistics relative to performance parameters for speed of task execution and task precision was made, to bring to the fore statistically significant differences between an expert surgeon and any novice (expert benchmarking) with equivalent number of simulator trials in the same set of camera view conditions. The individual task time and task precision statistics of one expert surgeon from a total of 120 training trials (single session) across 2D and 3D camera view conditions are shown here below in Tables 1a (time statistics) and 1b (precision statistics), together with the statistics of 10 novices (NT1-NT10) for the same number of training trials (single session) in the same camera view conditions, for comparison. The raw data on which these statistics were computed are provided in Table S1 of the Supplementary Materials Section. The statistics relative to task execution times (in seconds) of the expert did not differ much from those of eight of the ten novices in terms of means and their standard deviations. Two of the novices (NT5 and NT7) were considerably slower on average compared with the expert, with more variability around the means as indicated by the higher standard deviations (1a). The individual means for precision and its standard deviation show that the expert produced the smallest average off-target score (in pixels) with the least variability, indicating the highest level of task precision and performance stability for this criterion (1b). The ten novices were all considerably less precise with much more variability, as indicated by the higher average off-target scores, invariably two or more standard deviations higher than that of the expert, and the resulting, much larger, individual standard deviations around the means. The conclusion to be drawn from this analysis is that the experimental simulator task is perfectly capable of distinguishing the precision performance of a highly proficient surgical expert from the precision performance of surgical novices.

**Table 1a.** Expert session statistics for the parameter relative to **task time** (in seconds) compared with the session statistics of ten novice trainees (NT) for an equivalent number of individual simulator training trials (N = 120) across camera view conditions.

|  | EXPERT | NT 1 | NT2 | NT3 | NT4 | NT5 | NT6 | NT7 | NT8 | NT9 | NT10 |
|---|---|---|---|---|---|---|---|---|---|---|---|
| *Mean* | **13.74** | 15.79 | 14.79 | 12.90 | 14.81 | 26.23 | 19.17 | 21.76 | 13.46 | 12.46 | 12.82 |
| *Standard deviation* | **3.10** | 3.54 | 3.92 | 2.79 | 2.64 | 4.01 | 5.72 | 5.55 | 2.69 | 2.16 | 3.45 |

**Table 1b.** Expert session statistics for the parameter relative to **task precision** (off-target score in pixels) compared with the session statistics of ten novice trainees (NT) for an equivalent number of individual simulator training trials (N = 120) across camera view conditions.

|  | EXPERT | NT1 | NT2 | NT3 | NT4 | NT5 | NT6 | NT7 | NT8 | NT9 | NT10 |
|---|---|---|---|---|---|---|---|---|---|---|---|
| *Mean* | **871** | 2004 | 1598 | 1691 | 1189 | 1255 | 1229 | 1743 | 1425 | 1572 | 1919 |
| *Standard deviation* | **273** | 504 | 399 | 487 | 406 | 345 | 446 | 584 | 586 | 470 | 640 |

*3.2. Speed–Accuracy Trade-off Functions (SATFs) for Detecting Individual Strategies*

In the next analysis, the individual speed–precision trade-off functions of four novices who trained in a larger number of sessions across 2D camera view conditions (20 in the case of novice A, and eight in the cases of novices B, C, and D) in the same camera view conditions are shown. The motivation for SATFs is explained in detail in many statistics manuals and lecture notes, such as https://engineering.purdue.edu/~ece511/LectureNotes/pp19.pdf. The speed data are sorted in ascending order and plotted on the x-axis, with their corresponding precision scores on the y-axis. The raw data for these comparisons are made available in Table S2 of the Supplementary Materials section. The task execution times from a total of 1600 trials for novice A and a total of 640 trials for each of the other three novices (B, D, and C) are plotted in ascending order, together with their corresponding off-target scores (higher scores indicating lesser precision). These functions reveal, at a glance, the minima and maxima of the individual distributions for time and precision and their co-variability (scatter), and thereby highlight differences in individual training strategies (Figure 2).

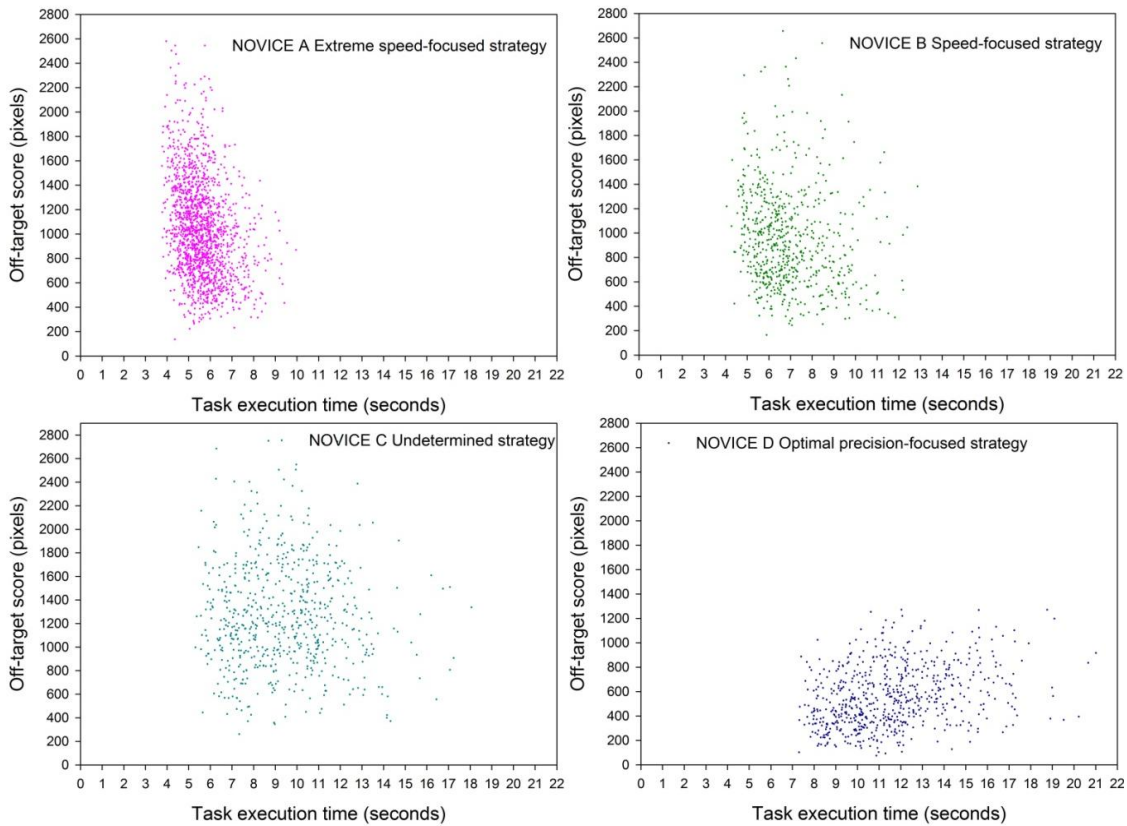

**Figure 2.** Individual trade-off functions between task execution times, plotted in ascending order, and their corresponding off-target scores (precision measure) for four novices who trained in a large number of sessions across 2D camera view conditions. The data reveal four different training strategies.

Novice A's function reveals an extreme speed-focused strategy, with time data between 4 and 10 s on the x-axis, to the detriment of the precision scores represented by off-target scores between 200 and 2600 pixels on the y-axis. Novice B's function reveals a less extreme, still mostly speed-focused, strategy with time data between 4 and 13 s on the x-axis, again to the detriment of the precision scores represented by off-target scores between 200 and 2700 pixels on the y-axis. Novice C's function is indicative of an undetermined strategy with no clear focus, with time data between 6 and 19 s on the x-axis and precision represented by off-target scores between 200 and 2700 pixels on the y-axis. Novice D's function reveals a precision-focused strategy with time data between 7 and 21 s on the x-axis and the highest precision, represented by off-target scores between 100 and 1200 pixels on the y-axis. The reasons why such strategy differences occur are not known, but it is made clear that starting off with a focus on the speed of task execution, which is a frequent strategy in untrained novices [2], is detrimental to improvement in precision, as shown here on the example of the training data of novices A and B. Undetermined strategies with no clear focus on either speed or precision (data of novice C) are not helping to improve performance either. In conclusion, the strategy that needs to be encouraged and selected for during training is definitely the one focused on precision (data of novice D), because it generates the highest precision scores even though performance may be a little slower on average, and bearing in mind that the purpose of surgical training is to aim for the highest level of precision, not the fastest task execution time.

This will be made even clearer by the next analysis, where the statistics from the last of ten (novice A) or eight (novices B, C, and D) training sessions of the four novice trainees are compared to the expert's single session statistics (same camera view conditions).

*3.3. Who Beats the Expert?*

Means and standard deviations for parameters relative to task time and task time precision are given in Tables 2a and 2b below, respectively. The raw data on which these statistics were computed are made available in Table S3 in the Supplementary Materials Section. Each individual statistic was computed on the basis of the same number of trials across 2D camera view conditions. These statistics reveal that the novices focused on speed (A and B) were considerably faster than the expert, as shown on the basis of their means for the time parameter and their standard deviations (2a), however, in the last one of their eight to ten training sessions they were nowhere near as precise as the expert after only a single session, as shown on the basis of their means for the precision parameter (off-target scores) and their standard deviations (2b).

**Table 2a.** Expert single session statistics for the parameter relative to **task time** (in seconds) compared with the statistics from the last of eight or ten training sessions of novice trainees A, B, C, and D in the same camera view conditions. Each statistic was computed on a total number of 80 trial sets.

|  | **EXPERT** | **NOVICE A** "Extreme Speed-Focused Strategy" | **NOVICE B** "Speed-Focused Strategy" | **NOVICE C** "Undetermined Strategy" | **NOVICE D** "Optimal Precision-Focused Strategy" |
|---|---|---|---|---|---|
| *Mean* | **14.63** | 4.76 | 6.35 | 8.85 | 9.13 |
| *Standard deviation* | **2.59** | 0.42 | 0.71 | 1.77 | 1.25 |

**Table 2b.** Expert single session statistics for the parameter relative to **task precision** (off-target score in pixels) compared with the statistics from the last of eight or ten training sessions of novice trainees A, B, C, and D in the same camera view conditions. Each statistic was computed on a total number of 80 trial sets.

|  | **EXPERT** | **NOVICE A** "Extreme Speed-Focused Strategy" | **NOVICE B** "Speed-Focused Strategy" | **NOVICE C** "Undetermined Strategy" | **NOVICE D** "OPTIMAL Precision-Focused Strategy" |
|---|---|---|---|---|---|
| *Mean* | **770** | 1146 | 905 | 1278 | 406 |
| *Standard deviation* | **166** | 378 | 250 | 434 | 151 |

The same conclusions hold for the novice with an undetermined speed–precision strategy (C). The only one of the four who, after eight training sessions, managed to "beat the expert" was the novice who adopted a precision-focused task strategy. The speed–precision function of this novice (Figure 2) shows markedly longer task times with greater scatter compared with that of the speed-focused novices, yet, in the last training session this novice had nonetheless become faster than the expert in a single session and, more importantly, also more precise. The conclusion to be drawn from these analyses is that speed–precision strategies of novices need to be detected and controlled for as early as possible in simulator training to ensure that trainees will attain optimal precision scores. The steps that are necessary for an automatic procedure to achieve this goal are conceptualized in the next section. As discussed in some of our previous work [4], the strategy differences between novices in simulator training for image-guided hand-tool movements generally vary between the two extreme cases (novices A and D) shown here, with a bias towards speed-focused strategies in absolute beginners. This calls for implementing systems which automatically detect, monitor, and if necessary correct the evolution of individual performance strategies, and provide the right kind of feedback to the trainee. Trainees who start off too fast need to be corrected to enable effective precision learning, while the performance strategy of trainees who focus on being as precise as possible should be reinforced, because they will naturally and without any further instruction also get faster with training. When an individual precision performance can be considered optimal and stable, then, and only then, the trainee may be instructed to try to get even faster. The performance profile of an expert should serve as a benchmark profile. This requires exploiting expert performance statistics for generating in-built system knowledge of what the desired performance profile of a novice should look like after successful training on a given simulator, with clear criteria for strategy, level of performance, and stability of performance.

*3.4. Criteria for Strategy*

Considering the analyses above, we propose a system that automatically records and stores performance data relative to time and precision in terms of cumulative values, that allows computing of session statistics: means, medians, and the minima and maxima for each individual trainee in a given trial set or training condition. These data can then be directly used to automatically plot the individual speed–accuracy functions for visualization. These functions should be made accessible and visible on the training screen as early as possible, ideally after the first session, and clearly identify the kind of strategy the trainee has followed.

*3.5. Criteria for Level of Performance*

Level of performance in terms of session means, medians, minima, and maxima of an individual trainee need to be compared against the benchmark statistics of an expert, ideally with a large number of simulator training sessions, to ensure the benchmarks capture the expert's full potential on the simulator for assessing relative levels of performance of individual trainees. As pointed out by others earlier [8], consensus-based criteria for "good" or "bad" performance make little sense in surgical training, as we need to train individuals to get as precise as possible at performing specific clinical trials at levels as close as possible to perfection. Therefore, using a surgical expert's level of performance as a training criterion in simulator tasks appears the ecologically most valid solution as it allows the simulator to be validated at the same time.

*3.6. Criteria for Stability of Performance*

Similarly, the stability of any performance measure is statistically defined and can, therefore, not be assessed in terms of any consensus-based criteria. A smaller standard deviation of the mean of a given performance measure indicates a smaller variability of the numerical data and, therefore, a more stable performance reflected by these data. This is well-illustrated by the analyses above, and makes a good case in favor of using expert benchmark statistics to automatically assess individual performance evolution. A true expert will by definition, since stability of performance is a characteristic of expertise, produce a stable performance with markedly smaller standard deviations around the mean of a given performance measure compared with any novice. Whenever this is not the case, then either the expert is not (or not yet) a true expert in the simulator task (no transfer of surgical expertise, for example), the simulator system as a whole is invalid, or the performance measure exploited reflects none of the specific aspects of surgical expertise

*3.7. Towards Expert-Based Speed–Precision Control*

In the light of what is considered above, a control system likely to enable trainees to reach an optimal speed–precision strategy as early as possible through automatically guided simulator training should be able to:

- generate reliable and discerning measures (parameters), relative to time and precision of individual performance, at any moment in time during training.
- compare individual parameter measures and statistics with the desired parameter value, based on the known ("learned") performance profile of an expert user, at any moment in time during training.
- provide feedback to the user as early as possible, and regularly as necessary, about what exactly he/she needs to focus on while training to attain an optimal performance level.

How this may be achieved is illustrated here by the example of the five step trial sequence of the simulator task (Figure 3). A single trial of the image guided pick-and-place task has several (here five, but it could be any $n$ in any other system) successive steps. The system starts counting task execution time from the moment the object is picked up by the user with the surgical tool ($t_0$), and ongoing time can be communicated to the trainee at any moment ($t_n$) from then until the object is placed on the last of several (here five) successive targets. Placing the object on a given target is a critical step of the procedure where precision matters, as users are instructed to place the object with the surgical tool on the central area of each target as precisely as possible. This is a challenging task and involves specific visual attention to fine eye-hand-tool coordination for placing the target optimally, as the precise borders of the target center are only known by the system in terms of pixel coordinates, but are not visible to the user. The user only sees the borders of the targets as such in the image guiding his/her action, and the object that needs to be placed centrally is smaller than the target area.

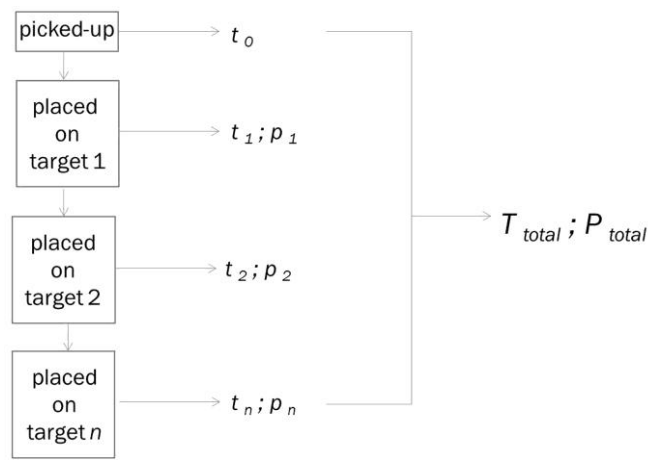

**Figure 3.** System flowchart of a single trial of the image guided pick-and-place task with $n$ successive critical "place object" steps. The system starts counting task execution time from the moment the object is picked up by the user with the surgical tool ($t_0$). Ongoing time can be communicated to the trainee at any moment ($t_n$) from then, together with the precision score ($p_n$), which is expressed here in terms of a "pixels-off-target score". The larger the precision measure $P$ as defined, the lesser the trainee's precision.

At each such critical step of the procedure, the system counts the number of pixels corresponding to the object in the image that do not coincide with the pixels that define the central area of a target known by the system, and are therefore "off-target" in terms of the task constraints as given ("place object as centrally on target as possible"). Hence, the smaller this measure (the off-target score), the greater the user's precision at a given critical step. The precision score ($p_n$) can also be communicated to the trainee at any critical moment in time ($t_n$) of the procedure. Thus, by automatically monitoring the evolution of individual performance parameters at any given moment in time during training, it is possible to control performance strategies of trainees and to ensure effective precision learning. It goes without saying that priority needs to be placed on precision rather than speed, especially in surgical training, and trainees get faster naturally, as shown above, once they have adopted the right strategy for working on their precision. As is shown here, a single dataset from a single expert can provide effective benchmark data for building prior knowledge into the system, and these "learned" data can be exploited for automatic performance feedback to the user at any moment in time during training. The data here from our expert were from a single session. In an ideal world, expert data could be collected from multiple simulator sessions, as many as necessary, to allow for matched trial-by-trial comparisons, where the observed data of a trainee at a given moment of the procedure for a given session $S_n$ are compared to the "perfect" data of an expert for the corresponding moment of the procedure and session $S_n$ of a training sequence. In the real world, the four cases (Figure 4) to be considered by a performance control system may be summarized as follows. (1) At a given moment $t_n$ in a training session $S_n$ the trainee is as fast as or faster than the expert and less precise. In this case, the system needs to alert him/her to slow down and start focusing on precision. This is the classic case of a trainee focused on speed who tries to do the task as fast as he/she can, and thereby compromises the swift evolution of his/her precision score. (2) At a given moment in a training session the trainee is slower than the expert and less precise. In this case, especially at early moments of training, the system needs to instruct the trainee to keep going, as he/she should get more precise and faster naturally. (3) At a given moment in a training session the trainee is slower than the expert and as precise or more precise. In this case, the system needs to instruct the trainee to try to go a little faster. (4) At a given moment the trainee is faster than or as fast as the expert and as precise or even more. In this case the trainee has beaten the expert. If this occurs, especially early in a training sequence, there is either a problem with the simulator task (i.e., the task does not produce adequate performance data that allow discriminating between levels of expertise, which is a problem that needs to be fixed), or the trainee is not a true novice.

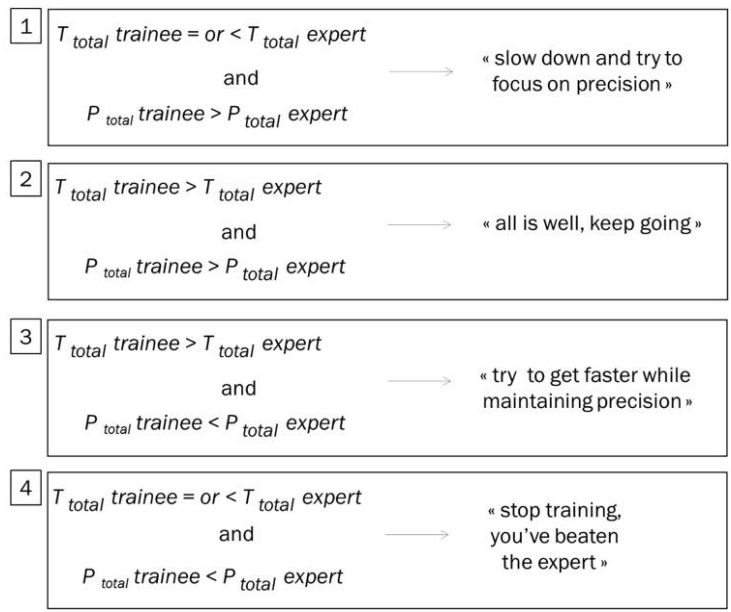

**Figure 4.** Decision model for the automatic correction or reinforcement of individual strategies during simulator training based on the expert's benchmark statistics ("learned" system knowledge). No more than four cases (1–4) need to be considered for generating clear performance feedback any novice can understand and take into account to attain an optimal speed–precision trade-off.

## 4. Discussion

Surgical simulator training requires new forms of sensorimotor learning, an adaptive process that leads to improvement in performance through practice. This adaptive process consists of multiple distinct learning processes [30-32]. Hitting a target or getting progressively closer to it generates by itself a form of implicit reward, where the trainee increasingly feels in control.

Successful error reduction, which is associated with specific commands relative to a motor task [26], can be optimized by giving a trainee the right external feedback. In this feedback process, the integration of information from multiple senses (vision, touch, audition, proprioception) leads to improved adjustments in body, arm, or hand movements. As a result, the task is performed with greater precision. Subjects are able to make good use of error signals relative to the discrepancy between a desired and the actual movement or hand-tool-position, or a discrepancy between visual and proprioceptive estimates of body, arm, or hand positions [24,30,31]. The effective, if possible computer controlled, monitoring of strategies relative to speed–accuracy trade-offs in individual performance learning is therefore a critical aspect of the skill assessment process.

Cognitive theories of motor learning predict that strategy differences occur spontaneously when novices train to perform a motor task in a limited number of sessions [24–28], as is indeed the case in laparoscopic simulator training. Conditional accuracy functions relate the duration of trial or task execution to a precision index reflecting the accuracy of the performance under conditions given, and changes in this relationship between speed and precision across sessions reflect hidden aspects of learning a beginner is usually not aware of [26,27]. In the fully trained expert, the trade-off between speed and precision does not vary markedly.

For a skill evaluator, the individual speed–accuracy trade-offs allow assessment of whether a trainee is indeed progressing or not. This knowledge needs to be made available as early as possible in the training process. Simply comparing the skill levels of different trainees at the end of the process is not the right approach. Objective (i.e., numerical) benchmark criteria for what the "ideal performance" of a successful trainee is to look like at the end of training are needed. Such benchmark knowledge can then be built into systems that automatically monitor the simulator task on the basis of results from a certain number of training sessions of a surgical expert, illustrated by our example above. Surgical simulators may be more or less specific to a physical model of surgical reality, i.e., "realistic", compared with the actual surgical task constraints they are supposed to train for. Some of them provide a variety of task specific feedback data, yet, the skills learnt on the given simulator may not transfer to other simulator tasks or physical models. One of the most important advantages of simulator training in the context of surgery is to facilitate skill evolution outside the clinical context, which reduces the risk for patients. Different definitions of the notion of skill itself have produced different approaches to simulation-based surgical training. However, as pointed out by others previously [11,12,20], it is not always clear if more skilled individuals perform better on their assessments than less skilled or experienced individuals (construct validity), whether individuals who perform well on their evaluations will also perform well on other similar or vaguely related tasks (concurrent validity), or whether an assessment based on simulator training will predict future performance in the real-world context (predictive validity).

Faced with this problem of providing reliable performance standards, it is essential that the system, the task, the metrics used to control performance learning during the task, and the mechanisms for providing feedback have somehow been validated by an expert to ensure that the training criteria and skill assessment provided by the system match those required for performing real surgical tasks. Since many different physical task models exist, surgical simulator training is permanently confronted with a problem of generalization of the learning curves and, ultimately, skill transfer to real-world surgical interventions.

The task model and control principles conceptualized in this work should be implemented at the earliest stages of "dry lab" simulator training. It could be adapted to a variety of eye-hand-tool coordination tasks that allow for computer controlled criteria relative to task precision $p$ at any critical task moment in time $t$. Early simulator training tasks should successfully tell apart the performance levels of a large number of novices from those of a surgical expert, not necessarily trained on the simulator but exhibiting a stable near-optimal performance with respect to task precision, as in the case discussed above. If an early training system satisfies this criterion, then it is indeed likely to measure critical aspects of surgical skill that will transfer to real surgical tasks. Ultimately, this should help produce a selection of trainees that will perform better later on, in more specific tasks on physical models, or in a clinical context. Then, direct supervision by experts will allow fostering individual expertise even further to produce excellence at the highest level of surgical proficiency (Figure 5).

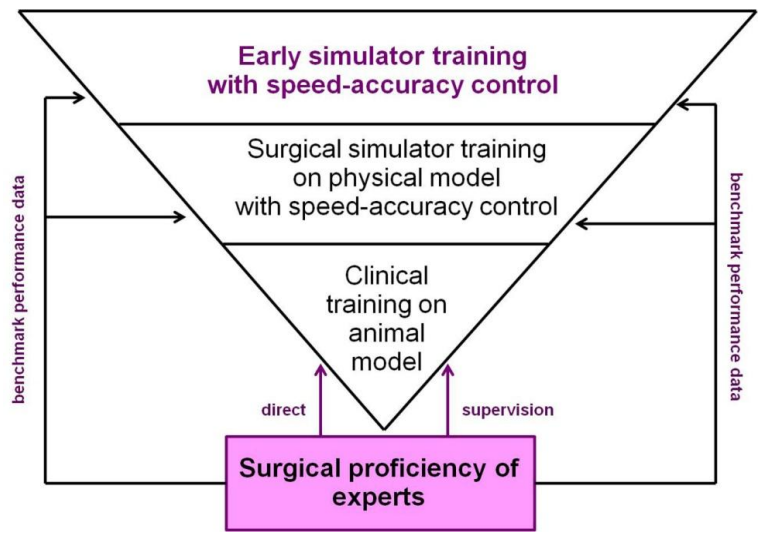

**Figure 5.** Speed–accuracy control on novice strategies in early training systems based on experts' benchmark performance data allows selection for critical aspects of surgical skill that will transfer to further training on physical models in increasingly realistic surgical tasks. Ultimately, this will produce a valid selection of trainees that are likely to perform well at the highest levels of training under the direct supervision of an expert.

Finally, whatever the simulator, a single performance metric inevitably gives a partial assessment of user performance [20]. Task completion time as a sole criterion has been explicitly demonstrated to be a poor or even misleading measure of surgical skill [1,23]. Some metrics assume a simple global optimum value, such as a minimal tool path length, or a minimal completion time, and other quantities such as forces [6,21] or velocities [23,33–35]. The ideal values of these may vary in relation to changes in conditions, which may have to be considered. Analysis of expert performance only will give insight into the nature of such dependencies and help to develop better simulators. The fact that some important elements of surgical proficiency have not, or not yet, been explored undeniably adds heuristic value to new conceptual approaches to unsupervised training, like the approach brought forward in this article. The latter can, in principle, be adapted to any simulator system that exploits criteria for task precision in limited task time. It is based on previously validated cognitive models of human performance learning that have shown that the individual speed–precision strategies of novices, which occur spontaneously and unconsciously [4,24–27], may compromise precision learning at all further stages of training if not controlled and corrected for as early as possible in the process. In robot-assisted surgical procedures, where the camera moves along with the tool, for example [10–16], metrics such as camera movement frequency, camera movement duration, or camera movement interval are important indicators of technical skill, i.e., the proficiency/precision with which the trainee controls the tool, in combination with other performance metrics such as task completion time, economy of tool motion, or master workspace range. An automatic training control procedure of the kind proposed in this article could be adapted to any such performance criteria, provided device-specific expert performance benchmarks are available.

**Funding:** This research received no external funding.

**Acknowledgments:** This research benefitted from material support from the Centre National de la Recherche Scientifique (CNRS) and the University of Strasbourg. The contribution of Anil Ufuk BATMAZ, in charge of programming the simulator system and running trainee surgeons in the context of his employment as a doctoral student with the University of Strasbourg, is acknowledged.

**Conflicts of Interest:** The author declares no conflict of interest.

## References

1. Stunt, J.J.; Wulms, P.H.; Kerkhoffs, G.M.; Dankelman, J.; van Dijk, C.N.; Tuijthof, G.J.M. How valid are commercially available medical simulators? *Adv. Med. Educ. Pract.* **2014**, *5*, 385–395.
2. Marcano, L.; Komulainen, T.; Haugen, F. Implementation of performance indicators for automatic assessment. In Proceedings of the 27th European Symposium on Computer Aided Process Engineering, Barcelona, Spain, 1–5 October 2017.
3. Marcano, L.; Yazidi, A.; Ferati, M.; Komulainen, T. Towards Effective Automatic Feedback for Simulator Training. In Proceedings of the 58th Conference on Simulation and Modelling, Reykjavik, Iceland, 25–27 September 2017.
4. Batmaz, A.U.; de Mathelin, M.; Dresp-Langley, B. Getting nowhere fast: Trade-off between speed and precision in training to execute image-guided hand-tool movements. *BMC Psychol.* **2016**, *4*, 55, doi:10.1186/s40359-016-0161-0.
5. Batmaz, A.U.; de Mathelin, M.; Dresp-Langley, B. Effects of indirect screen vision and tool-use on the time and precision of object positioning on real-world targets. *Perception* **2016**, *45*, 286.
6. Batmaz, A.U.; Falek, M.; Nageotte, F.; Zanne, P.; Zorn, L.; de Mathelin, M.; Dresp-Langley, B. Novice and expert haptic behaviours while using a robot controlled surgery system. In Proceedings of the 2017 13th IASTED International Conference on Biomedical Engineering, Innsbruck, Austria, 20–21 February 2017.
7. Batmaz, A.U.; de Mathelin, M.; Dresp-Langley, B. Seeing virtual while acting real: Visual display and strategy effects on the time and precision of eye-hand coordination. *PLoS ONE* **2017**, *12*, e0183789.
8. Batmaz, A.U.; de Mathelin, M.; Dresp-Langley, B. Effects of 2D and 3D image views on hand movement trajectories in the surgeon's peri-personal space in a computer controlled simulator environment. *Cogent Med.* **2018**, *5*, 1426232.
9. Gallagher, A.G.; O'Sullivan, C. *Fundamentals in Surgical Simulation: Principles and Practice. Improving Medical Outcome—Zero Tolerance Series*; Springer Business Media: Berlin, Germany, 2011.
10. Gallagher, A.G. Metric-based simulation training to proficiency in medical education: What it is and how to do it. *Ulster Med. J.* **2012**, *81*, 107–113.
11. Gallagher, A.G.; Ritter, E.M.; Champion, H.; Higgins, G.; Fried, M.P.; Moses, G. Virtual reality simulation for the operating room: Proficiency-based training as a paradigm shift in surgical skills training. *Ann. Surg.* **2005**, *241*, 364–372.
12. Dreyfus, H.L.; Dreyfus, S.E.; Athanasiou, T. *Mind over Machine: The Power of Human Intuition and Expertise in the Era of the Computer*; Free Press: New York, NY, USA, 1986.
13. Seymour, N.E.; Gallagher, A.G.; Roman, S.A.; O'Brien, M.K.; Andersen, D.K.; Satava, R.M. Analysis of errors in laparoscopic surgical procedures. *Surg. Endosc.* **2004**, *18*, 592–595.
14. Van Sickle, K.; Smith, B.; McClusky, D.A.; Baghai, M.; Smith, C.D.; Gallagher, A.G. Evaluation of a tensiometer to provide objective feedback in knot-tying performance. *Am. Surg.* **2005**, *71*, 1018–1023.
15. Van Sickle, K.R.; Gallagher, A.G.; Smith, C.D. The effect of escalating feedback on the acquisition of psychomotor skills for laparoscopy. *Surg. Endosc.* **2007**, *21*, 220–224.
16. Reznick, R.K. Teaching and testing technical skills. *Am. J. Surg.* **1993**, *165*, 358–361.
17. Chen, C.; White, L.; Kowalewski, T.; Aggarwal, R.; Lintott, C.; Comstock, B.; Kuksenok, K.; Aragon, C.; Holst, D.; Lendvay, T. Crowd-sourced assessment of technical skills: A novel method to evaluate surgical performance. *J. Surg. Res.* **2014**, *187*, 65–71.
18. Moorthy, K.; Munz, Y.; Sarker, S.K.; Darzi, A. Objective assessment of technical skills in surgery. *BMJ* **2003**, *327*, 1032–1037.
19. Sewell, C.; Morris, D.; Blevins, N.H.; Dutta, S.; Agrawal, S.; Barbagli, F.; Salisbury, K. Providing metrics and performance feedback in a surgical simulator. *Comput. Aided Surg.* **2008**, *13*, 63–81.
20. Ritter, E.M.; McClusky, D.A.; Gallagher, A.G.; Smith, C.K. Real-time objective assessment of knot quality with a portable tensiometer is superior to execution time for assessment of laparoscopic knot-tying performance. *Surg. Innov.* **2005**, *12*, 233–237.
21. Rosen, J.; Hannaford, B.; Richards, C.G.; Sinanan, M.N. Markov modeling of minimally invasive surgery based on tool/tissue interaction and force/torque signatures for evaluating surgical skills. *IEEE Trans. Biomed. Eng.* **2001**, *48*, 579–591.
22. Jarc, A.M.; Curet, M.J. Viewpoint matters: Objective performance metrics for surgeon endoscope control during robot-assisted surgery. *Surg. Endosc.* **2017**, *31*, 1192–1202.

23. Fogassi, L.; Gallese, V. Action as a binding key to multisensory integration. In *Handbook of Multisensory Processes*; Calvert, G., Spence, C., Stein, B.E., Eds.; MIT Press: Cambridge, MA, USA, 2004; pp. 425–441.
24. Bonnet, C.; Dresp, B. A fast procedure for studying conditional accuracy functions. *Behav. Res. Instrum. Comput.* **1993**, *25*, 2–8.
25. Fitts, P.M. The information capacity of the human motor system in controlling the amplitude of movement. *J. Exp. Psychol.* **1954**, *47*, 381–391.
26. Meyer, D.E.; Irwin, A.; Osman, A.M.; Kounios, J. The dynamics of cognition and action: Mental processes inferred from speed-accuracy decomposition. *Psychol. Rev.* **1988**, *95*, 183–237.
27. Luce, R.D. *Response Times: Their Role in Inferring Elementary Mental Organization*; Oxford University Press: New York, NY, USA, 1986.
28. Held, R. Visual-haptic mapping and the origin of crossmodal identity. *Optometry Vision Sci.* **2009**, *86*, 595–598.
29. Henriques, D.Y; Cressman, E. Visuo-motor adaptation and proprioceptive recalibration. *J. Motor Behav.* **2012**, *44*, 435–444.
30. Krakauer, J.W.; Mazzoni, P. Human sensorimotor learning: Adaptation, skill, and beyond. *Curr. Opin. Neurobiol.* **2011**, *21*, 636–644.
31. Goh, A.C.; Goldfarb, D.W.; Sander, J.C.; Miles, B.J.; Dunkin, B.J. Global evaluative assessment of robotic skills: Validation of a clinical assessment tool to measure robotic surgical skills. *J. Urol.* **2012**, *187*, 247–252.
32. Dresp-Langley, B. Principles of perceptual grouping: Implications for image-guided surgery. *Front. Psychol.* **2015**, 6, 1565.
33. Smith, R.; Patel, V.; Satava, R. Fundamentals of robotic surgery: A course of basic robotic surgery skills based upon a society consensus template of outcomes measures and curriculum development. *Int. J. Med. Robot Comput. Assist. Surg.* **2014**, *10*, 379–384.
34. Aiono, S.; Gilbert, J.M.; Soin, B.; Finlay, P.A.; Gordan, A. Controlled trial of the introduction of a robotic camera assistant (Endo Assist) for laparoscopic cholecystectomy. *Surg. Endosc. Other Interv. Tech*. **2002**, *16*, 1267–1270.
35. King, B.W.; Reisner, L.A.; Pandya, A.K.; Composto, A.M.; Ellis, R.D.; Klein, M.D. Towards an autonomous robot for camera control during laparoscopic surgery. *J. Laparoendosc. Adv. Surg. Tech.* **2013**, *23*, 1027–1030.